\documentclass[twocolumn,showkeys,preprintnumbers,amsmath,amssymb]{revtex4}
\usepackage{graphicx}% Include figure files
\usepackage{dcolumn}% Align table columns on decimal point
\usepackage{bm}% bold math

\begin{document}

\title{Magnetic behavior of PrPd$_{2}$B$_{2}$C}
\author{V. K. Anand,$^a$ A. Chaudhuri,$^a$ S. K. Dhar,$^b$ C. Geibel,$^c$ Z. Hossain$^a$}
%\email{vivekkranand@gmail.com}
\affiliation{$^a$Department of Physics, Indian Institute of Technology, Kanpur 208016, India}
\affiliation{$^b$Tata Institute of Fundamental Research, Mumbai-400005, India}
\affiliation{$^c$Max Planck Institute for Chemical Physics of Solids, 01187 Dresden, Germany}
%\email{ }
%\homepage{}
\begin{abstract}

We have synthesized a new quaternary borocarbide PrPd$_{2}$B$_{2}$C and measured its magnetization, electrical resistivity and specific heat. The compound crystallizes in the LuNi$_{2}$B$_{2}$C-type tetragonal structure (space group {\it I4/mmm}). Above 100 K the magnetic susceptibility follows Curie-Weiss behavior with effective moment $\mu_{eff}$ = 3.60 $\mu_{B}$, which is very close to the value expected for Pr$^{3+}$ ions. We do not find evidence for magnetic or superconducting transition down to 0.5 K. Specific heat exhibits a broad Schottky type anomaly with a peak at 24 K, very likely related to crystal electric field (CEF) excitation. The magnetic properties suggest the presence of a singlet CEF ground state leading to a Van-Vleck paramagnetic ground state.

\end{abstract}

%\pacs{75.47.De, 71.70.Ch, 71.27.+a, 75.30.Kz}% PACS, the Physics and Astronomy
%Use showkeys class option if keyword
\maketitle

%\section{\label{}Introduction\protect \lowercase{} }

\section*{Introduction}

The quaternary borocarbides RT$_{2}$B$_{2}$C (R = Sc, Y, Th, U or a rare earth, and T = Ni, Pt or Pd) have been investigated quite extensively due to their diverse physical properties like, e.g., superconductivity, magnetic ordering, coexistence of magnetism and superconductivity, heavy fermions and valence fluctuating behavior [1-4]. Among the Pr-based quaternary borocarbides, PrPt$_{2}$B$_{2}$C exhibits superconducting transition (T$_{c}$ $\sim$ 6 K) with nonmagnetic ground state [5,6] while PrNi$_{2}$B$_{2}$C orders antiferromagnectically (T$_{N}$ $\sim$ 4.3 K) with no signature of superconductivity down to 300 mK [7,8]. Here we report synthesis and physical properties of a new quaternary borocarbide PrPd$_{2}$B$_{2}$C. We also prepared the nonmagnetic analog LaPd$_{2}$B$_{2}$C. 

\section*{Experimental}

Polycrystalline samples of PrPd$_{2}$B$_{2}$C and LaPd$_{2}$B$_{2}$C were synthesized by the standard arc melting technique and annealed for one week at 900$^{o}$C. Samples were characterized by X-ray diffraction and Energy Dispersive X-ray Analysis (EDXA). Magnetization measurements were done using a SQUID magnetometer. Electrical resistance was measured using the standard ac four probe technique. Specific heat was measured using a commercial physical properties measurement system (Quantum Design).

\section*{Results and discussion}

The analysis of powder X-ray diffraction data reveal that both PrPd$_{2}$B$_{2}$C and LaPd$_{2}$B$_{2}$C form in the LuNi$_{2}$B$_{2}$C-type tetragonal structure (space group \textit{I4/mmm}). The lattice parameters \textit{a} and \textit{c} for PrPd$_{2}$B$_{2}$C and LaPd$_{2}$B$_{2}$C were found to be \textit{a} = 3.93 \AA, 3.95 \AA ~and \textit{c} = 10.44 \AA, 10.24 \AA, respectively.  They may be compared to CePd$_{2}$B$_{2}$C with \textit{a} = 3.912 \AA ~and \textit{c} = 10.277 \AA. From XRD and EDXA we estimate impurity phase(s) to be less than 5\% of the main phase.

\begin{figure}
\includegraphics[width=8.5cm, keepaspectratio]{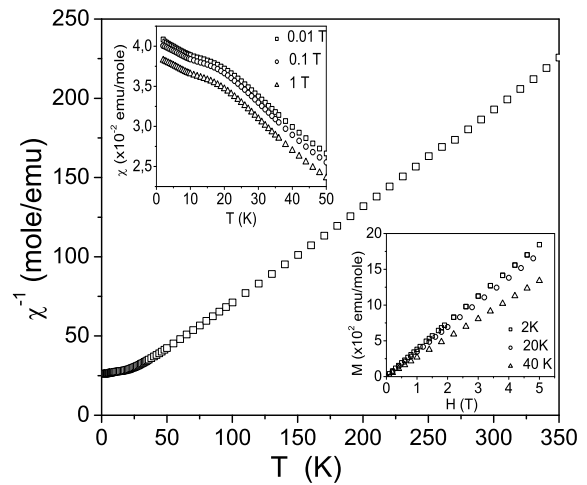}
\caption{\label{fig1} Inverse susceptibility of PrPd$_{2}$B$_{2}$C as a function of temperature measured in a field of 1.0 T. The insets show low temperature susceptibility at different fields and magnetization as a function of field.}
\end{figure}

Fig. 1 shows the magnetic susceptibility as a function of temperature. At a field of 1.0 T and above 100 K magnetic susceptibility follows a Curie-Weiss behavior. The effective moment $\mu_{eff}$ comes out to be 3.60 $\mu_{B}$, which is very close to the value expected for Pr$^{3+}$ ions (3.58 $\mu_{B}$). The paramagnetic Curie-Weiss temperature $\theta_{P}$ is found to be -14 K. Low temperature susceptibility data exhibits a weak shoulder at around 15 K which is present at all three fields of 0.01 T, 0.1 T and 1 T (shown in the upper inset of Fig. 1).  The isothermal magnetization (shown in the lower inset of Fig. 1) of PrPd$_{2}$B$_{2}$C at temperatures 2 K, 20 K and 40 K is linear with magnetic field.

\begin{figure}
\includegraphics[width=8.5cm, keepaspectratio]{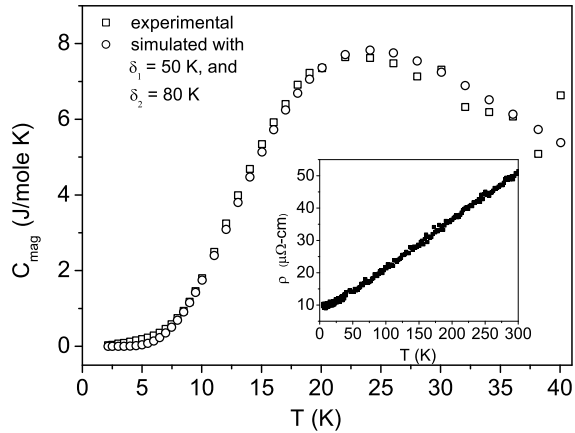}
\caption{\label{fig2} Magnetic heat capacity of PrPd$_{2}$B$_{2}$C as a function of temperature. Inset shows the temperature dependence of electrical resistivity.}
\end{figure}

Fig. 2 shows the magnetic contribution to the specific heat, obtained by subtracting off the nonmagnetic contribution from the total heat capacity of PrPd$_{2}$B$_{2}$C, assuming it to be approximately equal to the heat capacity of LaPd$_{2}$B$_{2}$C. A broad Schottky type anomaly with a peak around 24 K is observed. We do not see any pronounced anomaly in the specific heat of PrPd$_{2}$B$_{2}$C indicating the absence of magnetic transition. There is also no evidence of superconductivity down to 0.5 K. The experimentally observed magnetic heat capacity data could be reproduced well with a singlet ground state separated by first excited singlet at 50 K and a doublet at 80 K. 

The electrical resistivity (shown in the inset of Fig. 2) is found to decrease almost linearly with decreasing temperature and does not show any kind of anomaly down to 2 K. The residual resistivity $\rho_{0}$ = 10 $\mu \Omega$ cm and residual resistivity ratio (RRR) = 5.

\section*{Conclusions}

We synthesized a new quaternary borocarbide PrPd$_{2}$B$_{2}$C which does not exhibit any kind of phase transition down to 0.5 K. The magnetic properties suggest singlet CEF ground state with the first excited state at about 50 K. Since many of the borocarbides including LaPd$_{2}$B$_{2}$C and PrPt$_{2}$B$_{2}$C are superconducting, this compound provides us opportunity to investigate the interplay of superconductivity and Van-Vleck paramagnetism due to a singlet ground state. 

\vspace{0.5cm}
\textbf{Acknowledgments}

ZH acknowledges financial support from Indian Institute of Technology, Kanpur.  We thank S. Ghosh for his help in the initial stage of this work.

\end{document}